\documentstyle[12pt]{article}
\begin{document}
\thispagestyle{empty}
\newcommand{\be}{\begin{equation}}
\newcommand{\ee}{\end{equation}}
\newcommand{\sect}[1]{\setcounter{equation}{0}\section{#1}}
\newcommand{\vs}[1]{\rule[- #1 mm]{0mm}{#1 mm}}
\newcommand{\hs}[1]{\hspace{#1mm}}
\newcommand{\mb}[1]{\hs{5}\mbox{#1}\hs{5}}
\newcommand{\bea}{\begin{eqnarray}}
\newcommand{\eea}{\end{eqnarray}}
\newcommand{\wt}[1]{\widetilde{#1}}
\newcommand{\ux}[1]{\underline{#1}}
\newcommand{\ov}[1]{\overline{#1}}
\newcommand{\sm}[2]{\frac{\mbox{\footnotesize #1}\vs{-2}}
           {\vs{-2}\mbox{\footnotesize #2}}}
\newcommand{\prt}{\partial}
\newcommand{\eps}{\epsilon}
\newcommand{\p}[1]{(\ref{#1})}
\newcommand{\R}{\mbox{\rule{0.2mm}{2.8mm}\hspace{-1.5mm} R}}
\newcommand{\Z}{Z\hspace{-2mm}Z}
\newcommand{\cd}{{\cal D}}
\newcommand{\cg}{{\cal G}}
\newcommand{\ck}{{\cal K}}
\newcommand{\cw}{{\cal W}}
\newcommand{\vj}{\vec{J}}
\newcommand{\vl}{\vec{\lambda}}
\newcommand{\vz}{\vec{\sigma}}
\newcommand{\vt}{\vec{\tau}}
\newcommand{\vw}{\vec{W}}
\newcommand{\poiss}{\stackrel{\otimes}{,}}
\newcommand{\tx}{\theta_{12}}
\newcommand{\tb}{\overline{\theta}_{12}}
\newcommand{\zw}[1]{{#1 \over Z_{12}}}
\newcommand{\sqp}{{\alpha}}
\newcommand{\sqm}{{\overline\alpha}}
\newcommand{\nn}{\nonumber \\}

\newcommand{\NP}[1]{Nucl.\ Phys.\ {\bf #1}}
\newcommand{\PLB}[1]{Phys.\ Lett.\ {B \bf #1}}
\newcommand{\PLA}[1]{Phys.\ Lett.\ {A \bf #1}}
\newcommand{\NC}[1]{Nuovo Cimento {\bf #1}}
\newcommand{\CMP}[1]{Commun.\ Math.\ Phys.\ {\bf #1}}
\newcommand{\PR}[1]{Phys.\ Rev.\ {\bf #1}}
\newcommand{\PRL}[1]{Phys.\ Rev.\ Lett.\ {\bf #1}}
\newcommand{\MPL}[1]{Mod.\ Phys.\ Lett.\ {\bf #1}}
\newcommand{\BLMS}[1]{Bull.\ London Math.\ Soc.\ {\bf #1}}
\newcommand{\IJMP}[1]{Int.\ J.\ Mod.\ Phys.\ {\bf #1}}
\newcommand{\JMP}[1]{Jour.\ Math.\ Phys.\ {\bf #1}}
\newcommand{\LMP}[1]{Lett.\ Math.\ Phys.\ {\bf #1}}

\renewcommand{\thefootnote}{\fnsymbol{footnote}}
\newpage
\setcounter{page}{0}
\pagestyle{empty}
\vs{12}
\begin{center}
{\LARGE {\bf $N=4$ Sugawara construction on $\widehat{sl(2|1)}$,}}
\\ {\LARGE{\bf $\widehat{sl(3)}$ and mKdV-type superhierarchies
}}\\[0.8cm]

\vs{10} {\large E. Ivanov$^{a,1}$, S. Krivonos$^{a,2}$ and F.
Toppan$^{b,3}$} ~\\ \quad \\ {\em {~$~^{(a)}$ JINR-Bogoliubov
Laboratory of Theoretical Physics,}}\\ {\em 141980 Dubna, Moscow
Region, Russia}~\quad\\ {\em ~$~^{(b)}$ DCP-CBPF,}\\ {\em Rua
Xavier Sigaud 150, 22290-180, Urca, Rio de Janeiro, Brazil}
\end{center}
\vs{6}

\centerline{ {\bf Abstract}}

\vspace{0.3cm} \noindent The local Sugawara constructions of the
``small'' $N=4$ SCA in terms of supercurrents of $N=2$ extensions
of the affine $\widehat{sl(2|1)}$ and $\widehat{sl(3)}$ algebras are
investigated. The  associated super mKdV type hierarchies
induced by $N=4$ SKdV ones are defined. In the $\widehat{sl(3)}$ case the
existence of
two non-equivalent  Sugawara constructions is found. The ``long''
one involves
all the  affine $\widehat{sl(3)}$ currents, while the ``short'' one deals
only with  those from the subalgebra $\widehat{sl(2)\oplus u(1)}$. As a
consequence, the $\widehat{sl(3)}$-valued affine superfields carry two
non-equivalent mKdV type super hierarchies induced by the
correspondence between ``small'' $N=4$ SCA and $N=4$ SKdV
hierarchy. However, only the first hierarchy possesses genuine global
$N=4$ supersymmetry. We discuss peculiarities of the realization
of this $N=4$ supersymmetry on the affine supercurrents.

\vs{6} \vfill \rightline{CBPF-NF-046-99} \rightline{JINR E2-99-302}
\rightline{ solv-int/9912003} {\em E-Mail:\\ 1)
eivanov@thsun1.jinr.ru\\ 2) krivonos@thsun1.jinr.ru\\
3) toppan@cbpf.br}
\newpage
\pagestyle{plain}
\renewcommand{\thefootnote}{\arabic{footnote}}
\setcounter{footnote}{0}

\section{Introduction}

In the last several years integrable hierarchies of non-linear
differential equations have been intensely explored, mainly in
connection with the discretized two-dimensional gravity theories
(matrix models) \cite{DGZ} and, more recently, with the
$4$-dimensional super Yang-Mills theories in the
Seiberg-Witten approach \cite{SW}.

A vast literature is by now available on the construction and
classification of the hierarchies. In the bosonic case the understanding
of integrable hierarchies in $1+1$ dimensions is to a large extent
complete. Indeed, a generalized Drinfeld-Sokolov scheme \cite{DS}
is presumably capable to accommodate all known bosonic
hierarchies.

On the other hand, due to the presence of even and odd fields, the
situation for supersymmetric extensions remains in many respects
unclear. Since a fully general supersymmetric Drinfeld-Sokolov
approach to the superhierarchies is still lacking, up to now they
were constructed using all sorts of the available tools. These
include, e.g., direct methods, Lax operators of both scalar and
matrix type, bosonic as well as fermionic, coset construction,
etc. \cite{MRK}-\cite{Top}.

In \cite{IKT} a general Lie-algebraic framework for the $N=4$
super KdV hierarchy \cite{{DI},{DG2},{IK},{DG1}} and, hopefully, for its
hypothetical higher conformal spin counterparts (like $N=4$
Boussinesq) has been proposed. It is based upon a generalized
Sugawara construction on the $N=2$ superextended affine
(super)algebras which possess a hidden (nonlinearly realized)
$N=4$ supersymmetry. This subclass seemingly consists of $N=2$
affine superextensions of both the bosonic algebras with the
quaternionic structure listed in \cite{SSTP} and proper
superalgebras having such a structure. In its simplest version
\cite{IKT}, the $N=4$ Sugawara construction relates affine
supercurrents taking values in the $sl(2)\oplus u(1)$ algebra to
the ``minimal'' (or ``small'')  $N=4$ superconformal algebra
($N=4$ SCA) which provides the second Poisson structure for the
$N=4$ super KdV hierarchy. The Sugawara-type transformations are
Poisson maps, i.e. they preserve the Poisson-brackets structure of
the affine (super)fields. Therefore for any Sugawara
transformation which maps affine superfields, say, onto the
minimal $N=4$ SCA, the affine supercurrents themselves inherit an
integrable hierarchy which is constructed using the tower of the
$N=4$ SKdV hamiltonians in involution.
Such $N=4$ hierarchies realized on the affine supercurrents
can be interpreted as generalized mKdV-type superhierarchies.
The simplest example, the combined $N=4$ mKdV-NLS hierarchy
associated with the affine $N=2 \;\;\widehat{sl(2)\oplus u(1)}$
superalgebra, was explicitly constructed in \cite{IKT}.

In the case of higher-dimensional $N=4$ affine superalgebras this
sort of Sugawara construction is expected to yield additional
$N=4$ multiplets of currents which would form, together with those
of $N=4$ SCA (both ``minimal'' and ``large''), more general
nonlinear $N=4$ superalgebras of the $W$ algebra type.
Respectively, new SKdV (or super Boussinesq) type hierarchies with
these conformal superalgebras as the second Poisson structures can
exist, as well as their mKdV type counterparts associated with the
initial $N=4$ affine superalgebras. Besides, the linear $N=4$ SCAs
can be embedded into a given affine superalgebra in different
ways, giving rise to a few non-equivalent mKdV-type superhierarchies
associated with the same KdV-type superhierarchy.

In this paper we describe non-equivalent $N=4$ Sugawara
constructions for the eight-dimensional affine
(super)algebras $N=2 \;\;\widehat{sl(2|1)}$ and $N=2 \;\;\widehat{sl(3)}$.
These algebras are natural candidates for the higher-rank affine
superalgebras with hidden $N=4$ supersymmetry, next in complexity to
the simplest $\widehat{sl(2)\oplus u(1)}$ case treated in ref. \cite{IKT}.

The results can be summarized as follows.

In the $\widehat{sl(2|1)}$ case there are no
other {\em local} Sugawara constructions leading to the ``small''
$N=4$ SCA besides the one which proceeds from the
bosonic $\widehat{sl(2)\oplus u(1)}$ subalgebra supercurrents.
The $\widehat{sl(2|1)}$ affine
supercurrents carry a unique mKdV type hierarchy,
the evolution equations for the extra four superfields
being induced from their Poisson brackets with
the $N=4$ SKdV hamiltonians constructed from the
$sl(2)\oplus u(1)$-valued supercurrents.
The full hierarchy possesses by construction the
manifest $N=2$ supersymmetry and also reveals some extra exotic ``$N=2$
supersymmetry''. These two yield the standard
$N=4$ supersymmetry only on the $\widehat{sl(2)\oplus u(1)}$
subset of currents (``standard'' means closing on $z$
translations). Actually, such an extra $N=2$ supersymmetry
is present in {\it any}  $N=2$ affine (super)algebra with a
$\widehat{sl(2)\oplus u(1)}$ subalgebra. As the result,
neither the $N=2$ $\widehat{sl(2|1)}$ superalgebra itself, nor
the above-mentioned mKdV hierarchy reveal the genuine $N=4$
supersymmetry.

The $\widehat{sl(3)}$ case is more interesting since it admits
such an extended supersymmetry. In this case, besides the
``trivial'' $N=4$ SCA based on the $\widehat{sl(2)\oplus u(1)}$
subalgebra, one can define an extra $N=4$ SCA containing the full
$N=2$ stress-tensor and so involving all affine $\widehat{sl(3)}$
supercurrents \footnote{In what follows we name the corresponding
Sugawara construction  ``long'' $N=4$ Sugawara,  as opposed to the
``short'' one based on the $\widehat{sl(2)\oplus u(1)}$
subalgebra.}. We have explicitly checked that no other
non-equivalent local $N=4$ Sugawaras exist in this case. The
supercurrents of the second $N=4$ SCA  generate global $N=4$
supersymmetry closing in the standard  way on $z$-translations.
The defining relations of the $N=2$ $\widehat{sl(3)}$ algebra are
covariant under this supersymmetry, so it is actually $N=4$
extension of $\widehat{sl(3)}$, similarly to the
$\widehat{sl(2)\oplus u(1)}$ example. In the original  basis,
where the affine currents satisfy nonlinear constraints, the
hidden $N=2$ supersymmetry transformations are essentially
nonlinear and  mix all the currents. After passing, by means of a
non-local field redefinition, to the basis where the constraints
become  the linear chirality conditions, the supercurrents split
into some invariant  subspace and a complement which transforms
through itself and the invariant subspace. In other words, they
form a  not fully reducible representation of the $N=4$
supersymmetry. This  phenomenon was not previously encountered in
$N=4$ supersymmetric  integrable systems. We expect it to hold
also in higher rank  $N=2$ affine superalgebras with the hidden
$N=4$  structure.

The ``long'' Sugawara gives rise to a new mKdV type hierarchy
associated with the $N=4$ SKdV one. Thus the $\widehat{sl(3)}$
affine supercurrents provide an example of a Poisson structure
leading to two non-equivalent mKdV-type hierarchies, both associated
with $N=4$ SKdV, but recovered from the ``short'' and,
respectively, ``long'' $N=4$ Sugawara constructions. Only the
second hierarchy possesses global $N=4$ supersymmetry.

As a by-product, we notice the existence of another sort of super mKdV
hierarchies associated with both affine superalgebras considered.
They are related to the so-called ``quasi'' $N=4$ SKdV hierarchy 
\cite{{DGI},{DG2}}
which still possesses the ``small'' $N=4$ SCA as the second
Poisson structure but lacks global $N=4$ supersymmetry.
In the $\widehat{sl(3)}$ case there also exist two non-equivalent
``quasi'' super mKdV hierarchies generated through the ``short'' and
`long'' Sugawara constructions.

Like in \cite{IKT}, in the present paper we use the $N=2$ superfield
approach with the manifest linearly realized $N=2$ supersymmetry.
The results are presented in the language of
classical OPEs between $N=2$ supercurrents, which is equivalent
to the Poisson brackets formalism used in \cite{IKT}.
When evaluating these $N=2$ OPEs, we systematically exploit the
Mathematica package of ref. \cite{KT}.

\section{$N=2$ conventions and the minimal $N=4$ SCA}

Here we fix our notation and present the $N=2$ superfield
Poisson brackets structure of the ``minimal'' (``small'')
$N=4$ superconformal algebra (in the OPE language).

The $N=2$ superspace is parametrized by the coordinates
$Z\equiv \left\{ z, \theta , {\overline \theta}\right\}$,
with $\left\{ \theta , {\overline \theta} \right\}$
being Grassmann variables. The (anti)-chiral $N=2$
derivatives $D, {\overline D}$ are defined as
\begin{eqnarray}
D = \frac{\partial}{\partial \theta}
-\frac{1}{2}{\overline \theta} \partial_z \;,\;\;
{\overline D} = \frac{\partial}{\partial
{\overline \theta}} - \frac{1}{2}\theta\partial_z\; ,\;\;
D^2 = {\overline D}{}^2 = 0 \; , \;\;
\{ D, {\overline D}\} = -\partial_z \; \;. \label{Dcomm}
\end{eqnarray}

In the $N=2$ superfield notation the minimal $N=4$ SCA
is represented by the spin $1$ general superfield $J(Z)$ and two
(anti)-chiral spin $1$ superfields $W$, ${\ov W}$
($DW = {\ov D}\, {\ov W} =0$), with the following
classical OPE's
\begin{eqnarray}
{\ux {J(1)J(2)}} &=&  {2\over {Z_{12}}^2} - {{\tx \tb}\over {Z_{12}}^2}
J - \zw{\tb} {\ov D}J +\zw{ \tx} DJ -\zw{ \tx \tb} J' \;, \nonumber\\
{\ux {J(1)W(2)}} &=& -{{\tx\tb}\over {Z_{12}}^2} W - \zw{2} W -\zw{\tb} {\ov
D}W -\zw{\tx\tb} W' \;, \nonumber\\
{\ux {J(1){\ov W}(2)}} &=& -{{\tx\tb}\over {Z_{12}}^2} {\ov W} + \zw{2} {\ov W}
+\zw{\tx} D{\ov W} -\zw{ \tx\tb} {\ov W}' \;, \nonumber\\
{\ux {W(1){\ov W}(2)}} &=& {{\tx\tb}\over {Z_{12}}^3 } - {1\over {Z_{12}}^2}
-  {{\frac{1}{2} \tx\tb} \over {Z_{12}}^2} J  +\zw {\tb } {\ov D} J +\zw{1}
 J\; . \label{n4sca}
\end{eqnarray}
Here
$Z_{12} =
z_1 -z_2+\frac{1}{2}\left( \theta_1{\overline\theta}_2
-\theta_2{\overline\theta}_1\right)$, $\tx=\theta_1-\theta_2$, $\tb
={\overline\theta}_1-{\overline\theta}_2$, and the superfields
in the r.h.s. are evaluated at the point $(2)\equiv (\,z_2, \theta_2,
{\overline\theta}_2\,) $.

\section{The superaffinization of the $sl(2|1)$ superalgebra}

In this and next Sections we follow the general $N=2$ superfield
setting for $N=2$ extensions of affine (super)algebras \cite{HS,AIS}.

The $N=2$ $\widehat{sl(2|1)}$ superalgebra is generated by four fermionic
and four bosonic superfields, respectively
($H, {\overline H}, F, {\overline F}$) and ($S, {\overline S}, R,
{\overline R}$).
\par
The superfields $H, {\overline H}$ are associated with the Cartan
generators
of $sl(2|1)$ and satisfy the (anti)chiral constraints
\begin{eqnarray}
{\overline D}\, {\overline H} = D H = 0 \; \label{chir23}
\end{eqnarray}
while the remaining superfields are associated with the root
generators of $sl(2|1)$. In particular $F, {\overline F}$ are
related to the bosonic ($\pm$)-simple roots and, together with
$H, {\ov H}$, close on the superaffine ${\widehat {sl(2)\oplus u(1)}}$
subalgebra. The extra superfields satisfy the non-linear chiral
constraints
\begin{eqnarray}
&& {\overline D}\,{\overline R} =0 \;, \quad
{\overline D}\, {\overline F} = {\overline H}\,{\overline F} \; , \quad
{\overline D}\, {\overline S} = -{\overline F}\,{\overline R}
+{\overline H}\,{\overline S}\; , \nonumber \\
&& DR = HR \;, \quad DF = - H F\; ,
 \quad  DS = F R \;.\label{cond23}
\end{eqnarray}
The full set of OPEs defining the classical
$N=2$ superaffine ${\widehat{sl(2|1)}}$ algebra is given by
\begin{eqnarray}
&&{\underline{H(1){\overline H}(2)}} = {{\frac{1}{2}\tx\tb}\over {Z_{12}}^2}
- {1\over Z_{12}}\;, \;
{\underline{H(1)F(2)}} =  \zw{\tb} {F}\;, \;
{\underline{H(1){\overline F}(2)}}= - \zw{\tb} {\overline F}\; , \;
\nonumber\\
&&{\ux {H(1)S(2)}} = \zw{\tb} S\;,\;
{\ux {H(1) {\ov S}(2)}} = -\zw{\tb} {\ov S}\;, \;
{\ux {{\ov H} (1)F(2)}} = \zw{\tx} F\nonumber\; , \;
{\ux {{\ov H}(1){\ov F}(2)}} = -\zw{\tx} {\ov F}\; , \nonumber \\
&& {\ux {{\ov H}(1) R(2)}} = -\zw{\tx} R\; , \;
{\ux {{\ov H}(1){\ov R}(2)}} = \zw {\tx }{\ov R}\;, \nonumber\\
&& {\ux {F(1){\ov F}(2)}} = {{\frac{1}{2}\tx\tb}\over {Z_{12}}^2}
-\zw{1 -\tb {\ov H} - \tx H - \tx\tb
( F{\ov F} + H{\ov H} + {\ov D} H)} \;,
\nonumber\\
&&{\ux{F(1)S(2)}} = -\zw{\tx\tb} FS\;,\;
{\ux {F(1){\ov S}(2)}} 
= \zw{ \tb {\ov R} +\tx\tb (F {\ov S} +H{\ov R})}\;, \nonumber\\
&&{\ux {F(1){R}(2)}} = -\zw{\tb S + \tx\tb HS}\;,\;
{\ux {{\ov F} (1)S(2)}} = -\zw{\tx R + \tx\tb {\ov H} R}\;,\nonumber\\
&&{\ux {{\ov F}(1) R(2)}} = \zw{\tx\tb} R{\ov F}\;, \;
{\ux {{\ov F}(1) {\ov R}(2)}} =  \zw{ \tx {\ov S} -\tx\tb
({\ov F}\,{\ov R} -{\ov H}\,{\ov S})}\;,\nonumber\\
&&{\ux {S(1){\ov S}(2)}} = -{{\frac{1}{2}\tx\tb}\over {Z_{12}}^2}
+\zw{1 -\tb {\ov H} -\tx\tb (F{\ov F}- R{\ov R})} \;,\;
{\ux{S(1)R(2)}} = -\zw{\tx\tb} SR\;, \nonumber\\
&&{\ux {S(1){\ov R}(2)}} = \zw{\tx F +\tx\tb {\ov D}F}\;, \;
{\ux {{\ov S}(1)R(2)}} = \zw{\tb {\ov F}
+\tx\tb ( R{\ov S} + H {\ov F} - D{\ov F})}\;,\nonumber\\
&&{\ux {R(1){\ov R}(2)}} = -{{\frac{1}{2}\tx\tb}\over {Z_{12}}^2}
+ \zw{1 + \tx H +\tx\tb {\ov D} H} \;.\label{sope23}
\end{eqnarray}
All other OPEs are vanishing. The superfields
in the r.h.s. are evaluated at the point (2).

There is only one local Sugawara realization of $N=4$ SCA associated
with this affine $sl(2|1)$ superalgebra. It is explicitly given
by the relations
\begin{equation} \label{sl21N4}
J = {\ov D} H + D {\ov H} + H{\ov H} + F {\ov F}\; ,\;
W = D{\ov F}\; , \;
{\ov W} = {\ov D} F \; .
\end{equation}
It involves only the superfields ($H, {\ov H}, F, {\ov F}$)
which generate just the ${\widehat{sl(2)\oplus u(1)}}$-superaffine
subalgebra. It can be checked that no Sugawara construction
involving all the $sl(2|1)$ superfields exists in this case. The
$N=4$ SKdV hamiltonians constructed from the superfields \p{sl21N4}
produce an mKdV type hierarchy of the evolution equations for the
$\widehat{sl(2|1)}$ supercurrents through the OPE relations \p{sope23}.

Note that the supercurrents \p{sl21N4} generate
global non-linear automorphisms of $N=2$ $\widehat{sl(2|1)}$
(preserving both
the OPEs  \p{sope23} and the constraints \p{cond23}), such that their
algebra formally coincide with the $N=4$ supersymmetry algebra.
However, these
fermionic  transformations close in a standard way on $z$-translations
only on  the ${\widehat{sl(2)\oplus u(1)}}$ subset. On the rest of
affine supercurrents they yield complicated
composite objects in the closure. It is of course a consequence of
the fact that the true $z$-translations of all supercurrents are generated
by the full $N=2$ stress-tensor on the affine superalgebra, while
$N=4$ SCA \p{sl21N4} contains the stress-tensor on a subalgebra.
So this fermionic automorphisms symmetry cannot be viewed
as promoting the manifest $N=2$ supersymmetry
to $N=4$ one \footnote{This
kind of odd automorphisms is inherent to any $N=2$ affine algebra or
superalgebra  containing ${\widehat{sl(2)\oplus u(1)}}$ subalgebra.}.
Thus the $N=2$ superaffine $\widehat{sl(2|1)}$ algebra as a whole
possesses no hidden $N=4$ structure, as distinct from its
${\widehat{sl(2)\oplus u(1)}}$ subalgebra. This obviously implies
that the super mKdV hierarchy  induced on the full set of
the $\widehat{sl(2|1)}$ supercurrents  through the  Sugawara construction
\p{sl21N4} is not $N=4$ supersymmetric as well.

\section{The superaffine ${\widehat{sl(3)}}$ algebra}

The superaffinization of the $sl(3)$ algebra is spanned by eight
fermionic $N=2$ superfields subjected to non-linear (anti)chiral
constraints. We denote these superfields  $H, F, R, S$ (their antichiral
counterparts are ${\ov H}, {\ov F}, {\ov R}, {\ov S}$). The
${\widehat{sl(2)\oplus u(1)}}$ subalgebra is represented by $H, {\ov
H}, S, {\ov S}$. As before the Cartan subalgebra
is represented by the standard (anti)chiral $N=2$
superfields $H, {\overline H}$
\begin{eqnarray}
&& D H = {\ov D}\, {\ov H} = 0~. \label{chirH}
\end{eqnarray}
The remaining supercurrents are subject to
the non-linear constraints:
\begin{eqnarray}
&&DS = - HS\;, \quad D F = -\sqm H F + SR \;, \quad
     DR = \sqp H R \;, \nonumber \\
&&{\ov D}\, {\ov S} = {\ov H}\,{\ov S}\; ,\quad
{\ov D}\, {\ov F} = \sqp {\ov H}\,{\ov F} -{\ov S}\,{\ov R}\; , \quad
{\ov D}\,{\ov R} = - \sqm {\ov H}\,{\ov R}\;, \label{cond3}
\end{eqnarray}
where
\begin{equation} \label{param}
\sqp= \frac{1+i\sqrt{3}}{2} \;, \quad \sqm= \frac{1-i\sqrt{3}}{2} \;.
\end{equation}

The non-vanishing OPEs of the classical $N=2$ superaffine
${\widehat{sl(3)}}$ algebra read:
\begin{eqnarray}
&&{\ux{H(1) {\ov H}(2)}} = {{\frac{1}{2}\tx\tb}\over {Z_{12}}^2}- \zw{1} \;,\;
{\ux {H(1) F(2)}} = \zw{\sqp\tb} F \;,\;
{\ux {H(1) {\ov F}(2)}} = -\zw{\sqp \tb} {\ov F}\;, \nonumber\\
&&{\ux {H(1) S(2) }} = \zw{\tb} S\;, \;
{\ux {H(1) {\ov S}(2)}} = -\zw{\tb} {\ov S}\; ,\;
{\ux {H(1) R(2)}}= -\zw{\sqm \tb} R\; ,\;
{\ux {H(1) {\ov R}(2)}} = \zw{\sqm \tb} {\ov R}\; ,\nonumber\\
&&{\ux {{\ov H}(1) F(2)}} = \zw{\sqm  \tx} F\; , \;
{\ux {{\ov H}(1) {\ov F}(2)}} = -\zw{\sqm \tx} {\ov F}\; , \;
{\ux {{\ov H}(1) S(2)}} = \zw{ \tx} S\; , \;
{\ux {{\ov H}(1) {\ov S}(2)}} = - \zw{ \tx} {\ov S}\; , \nonumber\\
&&{\ux {{\ov H}(1) R(2)}} = -\zw{ \sqp \tx} R \; , \;
{\ux {{\ov H}(1) {\ov R} (2)}} = \zw{\sqp \tx} {\ov R}\; , \nonumber\\
&&{\ux { F(1){\ov F}(2)}} = {{\frac{1}{2}\tx\tb}\over {z_{12}}^2}
-\zw{1 - \sqp \tb {\ov H} -\sqm \tx H -
  \tx \tb
( F {\ov F} + H {\ov H} + R {\ov R}+ S{\ov S} +\sqm {\ov D} H)}\;,
\nonumber\\
&&{\ux {F(1) S(2) }} = \zw{\sqp \tx \tb} FS \; , \;
{\ux {F(1){\ov S}(2)}}= \zw{ \tx R + \tx \tb ({\ov D} R +
\sqm F{\ov S} - {\ov H} R )}\; ,\nonumber\\
&&{\ux { F(1)R(2)}} = \zw{ \sqm \tx\tb} FR\; , \;
{\ux {F(1){\ov R}(2)}} = -\zw {\tx S +\tx\tb ( {\ov D} S -\sqp F{\ov R}
+\sqm {\ov H} S)}\; ,\nonumber\\
&&{\ux { {\ov F}(1) S(2)}} = -\zw{ \tb {\ov R } - \tx \tb
( H{\ov R} -\sqp {\ov F} S + D {\ov R} )} \; , \;
{\ux { {\ov F}(1) {\ov S}(2)}} = -\zw{\sqm  \tx\tb}
{\ov F}\,{\ov S}\; , \nonumber\\
&&{\ux { {\ov F} (1) R(2)}} = \zw{ \tb {\ov S} - \tx \tb( D {\ov S}-
 \sqp H {\ov S} +\sqm {\ov F} R)} \; , \;
{\ux { {\ov F} (1) {\ov R} (2)}} 
= -\zw{ \sqp\tx \tb} {\ov F} \,{\ov R}\;, \nonumber\\
&& {\ux {S(1) {\ov S} (2)}} = {{\frac{1}{2}\tx\tb}\over{Z_{12}}^2}  -
\zw{1-  \tb {\ov H} - \tx H -  \tx \tb
(S{\ov S} + H {\ov H} + {\ov D} H )}\;,\; \nonumber\\
&&{\ux { S(1) R(2)}} = -\zw{ \tb F + \tx \tb (H F - \sqm S R )} \;, \;
{\ux {S(1){\ov R}(2)}} = -\zw{\sqm \tx\tb}  S{\ov R} \; ,\nonumber\\
&&{\ux{{\ov S}(1) R(2)}} = \zw{\sqp \tx\tb} {\ov S} R\; , \;
{\ux{{\ov S}(1){\ov R}(2)}} = \zw{ \tx {\ov F} + \tx \tb ({\ov H}\,
{\ov F} -\sqp {\ov S}\, {\ov R})}\; ,\nonumber\\
&&{\ux { R(1){\ov R}(2)}} = 
{{\frac{1}{2}\tx\tb}\over {Z_{12}}^2} -\zw{1 + \sqm \tb
 {\ov H} +\sqp \tx H - \tx\tb ( H{\ov H} + R {\ov R}
-\sqp {\ov D} H)} \;. \label{sl3}
\end{eqnarray}

There exist two non-equivalent ways to embed the affine supercurrents
into the minimal $N=4$ SCA via a local Sugawara construction. One
realization, like in the $\widehat{sl(2|1)}$ case, corresponds to the
``short'' Sugawara construction based solely upon the
$\widehat{sl(2)\oplus u(1)}$
subalgebra. The second one,
which in what follows is referred to as the ``long'' Sugawara
construction, involves {\it all} the $sl(3)$-valued affine supercurrents.
This realization corresponds to a new globally $N=4$
supersymmetric hierarchy realized on the full set of superaffine
$\widehat{sl(3)}$ supercurrents. Thus the set of superfields generating the
superaffine ${\widehat{sl(3)}}$ algebra supplies the first known example of a
Poisson-brackets structure carrying two non-equivalent hierarchies of the
super mKdV type associated with $N=4$ SKdV hierarchy.

The two Sugawara realizations are respectively given by:

{\em i)} in the ``short'' case,
\begin{equation} \label{short}
J =  D{\ov H} + {\ov D} H + H {\ov H} + S {\ov S}\; ,\quad
W = D {\ov S}\; , \quad
{\ov W} = {\ov D}S \;,
\end{equation}

{\em ii)} in the ``long'' case
\begin{equation} \label{long}
J=  H{\ov H} + F {\ov F} + R {\ov R} + S {\ov S}
+ \sqm{\ov D} H +\sqp D{\ov H}\; , \quad
W =  D {\ov F}\; , \quad
{\ov W} = {\ov D}F \;.
\end{equation}

Their Poisson brackets (OPEs) are given by the relations (\ref{n4sca}).

\section{$N=4$ supersymmetry}

Like in the $\widehat{sl(2\vert 1)}$ case, the ``short'' Sugawara
$N=4$ supercurrents  \p{short} do not produce the true global
$N=4$ supersymmetry for the entire set of the affine
supercurrents, yielding it only for the $\widehat{sl(2)\oplus
u(1)}$ subset. At the same time, the ``long'' Sugawara \p{long}
generates such a supersymmetry. In the $z, \theta, \bar{\theta} $
expansion of the supercurrents $J, W, {\ov W}$  the global
supersymmetry generators are present as the coefficients of the
monomials $\sim \theta / z$. {}From $J$  there come out the
generators of the manifest linearly realized $N=2$  supersymmetry,
while those of the hidden $N=2$ supersymmetry appear from $W, {\ov
W}$. The precise form of the hidden supersymmetry transformations
can then be easily read off from the OPEs \p{sl3}:
\begin{eqnarray}
\delta H & =  & {\ov\epsilon} \left( HF -\sqp \, SR \right)
      + \epsilon \sqp  \, D{\ov F} \; , \qquad
\delta {\ov H}  =  {\ov\epsilon}\, \sqm \, {\ov D} F
    -\epsilon \left( {\ov H}\,{\ov F} -\sqm \,{\ov S}\,{\ov R}
      \right) \;,  \nn
\delta F &=& -\epsilon\left( \sqp D{\ov H}+F{\ov F}+H{\ov H}+
     R{\ov R}+S{\ov S}\right)\,,
\delta{\ov F} = -{\ov \epsilon}\left(
   \sqm\,{\ov D}H+F{\ov F}+H{\ov H}+
     R{\ov R}+S{\ov S}\right)\;, \nn
\delta S & =  & -{\ov \epsilon} \sqp\, FS-\epsilon\left(
   D{\ov R}-\sqp\,{\ov F}S+H{\ov R}\right)\,, \;
\delta {\ov S} = -{\ov \epsilon} \left(
   {\ov D} R +\sqm \, F{\ov S}-{\ov H}R \right)
        +\epsilon \sqm\, {\ov F}\,{\ov S}\;, \nn
\delta R & = & -{\ov \epsilon}\,\sqm\,FR + \epsilon\left( D{\ov S}
     +   \sqm\,{\ov F}R -\alpha \, H{\ov S}\right)\,,\;
\delta {\ov R} =  {\ov \epsilon}\left( {\ov D} S-
  \sqp \, F{\ov R} +\sqm\,{\ov H} S \right)
  + \epsilon\alpha \,{\ov F}\,{\ov R} \;. \label{hidN2}
\end{eqnarray}
Here $\epsilon, {\ov \epsilon}$ are the corresponding odd
transformation parameters. One can check that these
transformations have just the same standard closure in  terms of
$\partial_z $ as the manifest $N=2$ supersymmetry
transformations, despite the presence of nonlinear terms. Also it
is  straightforward to verify that the constraints \p{cond3} and
the  OPEs \p{sl3} are covariant under these transformations.

Let us examine the issue of reducibility of the set of the $N=2$
$\widehat{sl(3)}$ currents with respect to the full $N=4$ supersymmetry.
In the $\widehat{sl(2)\oplus u(1)}$ case the involved currents form an
irreducible $N=4$ multiplet  which is a nonlinear version of the
multiplet consisting of two chiral (and anti-chiral) $N=2$
superfields \cite{IK}. In the given case  one can expect that eight $N=2$
$\widehat{sl(3)}$ currents form a reducible multiplet which can be divided
into  a sum of two irreducible ones, each involving  four superfields
(a pair of chiral and anti-chiral superfields together with
its conjugate). However, looking at the r.h.s. of \p{hidN2}, it is
difficult to imagine how  this could be done in a purely algebraic
and local way. Nevertheless, there is a non-local redefinition of the
supercurrents which partly makes this job. As the first step one
introduces a prepotential for the chiral superfields $H, {\ov H}$
\bea
H = DV, \quad {\ov H} = -{\ov D}\,{\ov V}
\eea
and chooses a gauge for $V$ in which it is expressed
through $H, {\ov H}$ \cite{Egau}
\bea
V &=& - \partial^{-1}({\ov D} H + \sqm \, D{\ov H})\;,\quad
{\ov V} =  \partial^{-1}(D {\ov H} + \sqp \, {\ov D}H) \;, \quad
V = - \sqm {\ov V}\;,  \label{relV} \\
\delta V &=& \sqp (\bar \epsilon F -\epsilon {\ov F}) \;, \qquad
\delta {\ov V} = \sqm (\bar \epsilon F -\epsilon {\ov F})
\;. \label{tranV}
\eea
Using this newly introduced quantity, one can pass to the
supercurrents which satisfy the standard chirality conditions
following from the original constraints \p{chirH}, \p{cond3}
and equivalent to them
\bea  S &=& \exp\{-V\}\tilde{S}\;, \quad {\ov S} =
\exp\{\sqp V\}{\ov {\tilde{S}}}
\;, \quad
R = \exp\{\sqp V\}\tilde{R}\;, \quad {\ov R} = \exp\{-V\}{\ov {\tilde{R}}}\;,
\nonumber \\
F &=& \exp\{-\sqm V\}[\tilde{F} -
\partial^{-1}{\ov D}(\tilde{S}\tilde{R}) + \partial^{-1}D
({\ov {\tilde{S}}}\,{\ov {\tilde{R}}})]\;,\nn
{\ov F} &=& \exp\{-\sqm V\}[{\ov {\tilde{F}}} -
\partial^{-1}{\ov D}(\tilde{S}\tilde{R}) + \partial^{-1}D
({\ov {\tilde{S}}}\,{\ov {\tilde{R}}})]\;,
\label{redef2}
\eea
\bea
D\tilde{S} = D \tilde{R} = D \tilde{F} = 0\;, \qquad
{\ov D}{\ov {\tilde{S}}} = {\ov D}{\ov {\tilde{R}}} =
{\ov D}{\ov {\tilde{F}}} =
0\;. \label{chirSRF}
\eea
The $N=4$ transformation rules \p{hidN2} are radically simplified in
the new basis
\bea
&& \delta \tilde{S} = -\epsilon D {\ov {\tilde{R}}} \;,\quad
\delta {\ov {\tilde{S}}} = -\bar\epsilon {\ov D} \tilde{R} \;,\quad
\delta \tilde{R} = \epsilon D {\ov {\tilde{S}}} \;,\quad
\delta {\ov {\tilde{R}}} = \bar\epsilon {\ov D} \tilde{S} \;, \nn
&& \delta \tilde{F} = \epsilon D {\ov D} (\exp\{\sqm V\})\;, \qquad
\delta {\ov {\tilde{F}}} = -\bar\epsilon {\ov D}D
(\exp\{\sqm V\})\;, \nn
&& \delta (\exp\{\sqm V\}) = \bar\epsilon \tilde{F}  -\epsilon {\ov
{\tilde{F}}}    -(\bar\epsilon - \epsilon)
\partial^{-1}[{\ov D}(\tilde{S}\tilde{R}) - D
({\ov {\tilde{S}}}\,{\ov {\tilde{R}}})] \;.\label{tranVn} \eea We
see that the supercurrents $\tilde{S}\;, {\ov {\tilde{S}}}\;,
\tilde{R}\;, {\ov {\tilde{R}}}$ form an irreducible $N=4$
supermultiplet, just of the kind found in \cite{IK}. At the same
time, the superfields $V, \tilde{F}\;, {\ov {\tilde{F}}}$ do not
form a closed set: they transform through the former multiplet. We
did not succeed in finding the basis where these two sets of
transformations entirely decouple from each other. So in the
present case we are facing a new phenomenon consisting in that the
$N=2 \;\;\widehat{sl(3)}$ supercurrents form a not fully reducible
representation of $N=4$ supersymmetry. The same can be anticipated
for higher rank affine supergroups  with a hidden $N=4$ structure.
One observes that putting the supercurrents $\tilde{S}\;, {\ov
{\tilde{S}}}\;, \tilde{R}\;, {\ov {\tilde{R}}}$ (or their
counterparts in the original basis) equal to zero is the
truncation consistent with $N=4$ supersymmetry. After this
truncation the remaining supercurrents $H,F, {\ov H}, {\ov F}$ form
just the same irreducible multiplet as in the
$\widehat{sl(2)\oplus u(1)}$ case \cite{IKT}.

Note that the above peculiarity does not show up at the level of
the composite supermultiplets like \p{long}. Indeed, it is straightforward to
see that the supercurrents in \p{long} form the same irreducible
representation as in the $\widehat{sl(2)\oplus u(1)}$ case \cite{IKT}
\bea
\delta J = -\epsilon {\ov D} W - \bar \epsilon D{\ov W}\;, \qquad
\delta W = \bar \epsilon D J\;, \qquad
\delta {\ov W} = \epsilon {\ov D} J\;. \label{JWtran}
\eea
Another irreducible multiplet is comprised by the following
composite supercurrents
\bea
\hat{J} &=& H{\ov H} + F{\ov F} + S{\ov S} +R{\ov R}\;, \nn
\quad  \hat{W} &=& DF = -\sqm HF +SR \;, \;  \hat{{\ov W}} =
{\ov D}\,{\ov F} = \sqp {\ov H}\,{\ov F} - {\ov S}\,{\ov R}~.
\label{top}
\eea
Under \p{hidN2} they transform as
\bea
\delta \hat{J} = -\epsilon D \hat{{\ov W}} -
\bar \epsilon {\ov D} \hat{W}\;, \qquad
\delta \hat{W} = \epsilon D \hat{J}\;, \qquad
\delta \hat{{\ov W}} = \bar\epsilon {\ov D} \hat{J}\;.
\label{JWtran2}
\eea
The OPEs of these supercurrents can be checked to generate another
``small'' $N=4$ SCA with zero central charge, i.e. a topological
``small'' $N=4$ SCA. The same SCA was found  in the
$\widehat{sl(2)\oplus u(1)}$ case \cite{IKT}. This SCA
and the first one together close on the ``large'' $N=4$ SCA in some particular
realization \cite{RASS,IKT}. Thus the
$N=2 \;\;\widehat{sl(3)}$ affine
superalgebra  provides a Sugawara type construction for this
extended SCA as well. It would be of interest to inquire whether this
superalgebra conceals  in its enveloping algebra any other SCA containing
$N=4$ SCA as  a subalgebra, e.g., possible $N=4$ extensions of nonlinear
$W_n$ algebras.

\section{$N=4$ mKdV-type hierarchies}

Both two non-equivalent $N=4$ Sugawara constructions, eqs.
({\ref{short}) and (\ref{long}), define Poisson maps. As a
consequence, the superaffine $sl(3)$-valued supercurrents inherit
all the integrable hierarchies associated with $N=4$ SCA.

The first known example of hierarchy with $N=4$ SCA
as the Poisson structure is $N=4$ SKdV
hierarchy (see \cite{DI}). The densities of the lowest hamiltonians from
an infinite sequence of the corresponding superfield hamiltonians
in involution, up to an overall normalization factor, read
\begin{eqnarray}
{\cal H}_1 &=& J\nonumber\\ {\cal H}_2 &=& -{1\over 2}( J^2 - 2 W{\ov
W})\nonumber \\
\relax {\cal H}_3 &=&
{1\over 2} (J [ D, {\overline D}]  J + 2 W {\overline W}' +{2\over 3} J^3
- 4 J W{\ov W})~.
\end{eqnarray}
Here the $N=2$ superfields $J$, $W$, ${\ov W}$ satisfy the
Poisson brackets (\ref{n4sca}).

Let us concisely denote by $\Phi_a$, $a=1,2,...,8$,
the $\widehat{sl(3)}$-valued superfields $H,F,R,S$ together with the barred
ones. Their evolution equations which, by construction, are
compatible with the $N=4$ SKdV flows, for the $k$-th flow
($k=1,2,...$) are written as
\begin{eqnarray}
\relax
\frac{\partial}{\partial t_k}\Phi_a (X,t_k) &=& \{ \int dY
{\cal H}_k (Y, t_k) , \Phi_a (X,t_k)\}~.
\end{eqnarray}
The Poisson bracket here is given by the superaffine $\widehat{sl(3)}$
structure (\ref{sl3}), with $X, Y$ being two different ``points'' of
$N=2$ superspace.

The identification of the superfields $J$, $W$, $
{\ov W}$ in terms of the affine supercurrents
can be made either via eqs. (\ref{short}), i.e. the
``short'' Sugawara, or via eqs. (\ref{long}),
that is the ``long'' Sugawara. Thus the same $N=4$ SKdV
hierarchy proves to produce two non-equivalent
mKdV type hierarchies for the affine supercurrents, depending on
the choice of the underlying Sugawara construction.
The first hierarchy is $N=2$ supersymmetric, while the other one
gives a new example of globally $N=4$ supersymmetric hierarchy.

Let us briefly outline the characteristic features of
these two hierarchies.

It is easy to see that
for the superfields $H, {\ov H}, S, {\ov S}$ corresponding to the
superaffine algebra $\widehat{sl(2)\oplus u(1)}$ as
a subalgebra in $\widehat{sl(3)}$,  the ``short'' hierarchy coincides
with $N=4$ NLS-mKdV hierarchy of ref. \cite{IKT}.
For the remaining $\widehat{sl(3)}$ supercurrents one gets
the evolution equations in the ``background'' of the
basic superfields just mentioned.

New features are revealed while examining the ``long'', i.e. $N=4$
supersymmetric $\widehat{sl(3)}$ mKdV hierarchy. It can be easily
checked that for all non-trivial flows $(k \geq 2)$ the evolution
equations for any given superfield $\Phi_a$ necessarily contain in the
r.h.s. the whole set of eight $\widehat{sl(3)}$ supercurrents.
In this case the previous $N=4$ NLS-mKdV hierarchy can also be
recovered. However, it is obtained in a less trivial
way. Namely, it is produced only after coseting out
the superfields $R, S$ and ${\ov R}, {\ov S}$, i.e. those associated
with the simple roots of $sl(3)$ (as usual, the passing to the
Dirac brackets is required in this case). As was mentioned in the
preceding Section, this truncation preserves the global $N=4$
supersymmetry.

Let us also remark that, besides the two mKdV
hierarchies carried by the superaffine $\widehat{sl(3)}$ algebra
and discussed so far, this Poisson bracket structure also carries
at least one extra pair of non-equivalent hierarchies of the mKdV type
possessing only global $N=2$ supersymmetry.
It was shown in \cite{DGI} (see also \cite{DG2}) that the enveloping 
algebra of $N=4$ SCA
contains, apart from an infinite abelian subalgebra corresponding
to the genuine $N=4$ SKdV hierarchy, also an infinite abelian
subalgebra formed by the hamiltonians in involution associated
with a different hierarchy referred to as the ``quasi'' $N=4$ SKdV one.
This hierarchy admits only a global $N=2$ supersymmetry
and can be thought of as an integrable extension of the $a=-2$, $N=2$ SKdV
hierarchy. In \cite{DGI} there was explicitly found a non-polynomial
Miura-type transformation which in a surprising way  relates $N=4$ SCA
to the non-linear $N=2$ super-$W_3$ algebra. This transformation
maps the ``quasi'' $N=4$ SKdV hierarchy onto the $\alpha=-2$, $N=2$
Boussinesq hierarchy. Since these results can be
rephrased in terms of the Poisson brackets structure alone, and the
same is true both for our ``short'' (\ref{short}) and ``long''
(\ref{long}) Sugawara constructions, it immediately follows that the
super-affine $\widehat{sl(3)}$ superfields also carry two non-equivalent
``quasi'' $N=4$ SKdV structures and can be mapped in two non-equivalent
ways onto the $\alpha=-2$, $N=2$ Boussinesq hierarchy.

\section{Conclusions}
In this work we have investigated the local Sugawara
constructions leading to the $N=4$ SCA expressed in terms of the
superfields corresponding to the $N=2$ superaffinization of the
$sl(2|1)$ and the $sl(3)$ algebras. We have shown that the
$\widehat{sl(3)}$ case admits a non-trivial $N=4$ Sugawara
construction involving all eight affine supercurrents and
generating the hidden $N=4$ supersymmetry of $N=2
\;\;\widehat{sl(3)}$ algebra. This property has been used to
construct a new $N=4$ supersymmetric mKdV hierarchy associated
with $N=4$ SKdV. Another mKdV hierarchy is obtained using the
$N=4$ Sugawara construction on the subalgebra
$\widehat{sl(2)\oplus u(1)}$. Thus the $N=2$ $\widehat{sl(3)}$
algebra was shown to provide the first example of a Poisson
brackets structure carrying two non-equivalent integrable mKdV type
hierarchies associated with the $N=4$ SKdV one. Also, the
existence of two non-trivial $N=2$ supersymmetric mKdV-type
hierarchies associated with the same superaffine Poisson structure
and ``squaring'' to the quasi $N=4$ SKdV hierarchy of ref.
\cite{DGI} was noticed.

An interesting problem is to generalize the two Sugawara constructions
to the full quantum case and to find out (if existing) an $N=4$ analog
of the well-known GKO coset construction \cite{GKO} widely used in
the case of bosonic affine algebras. It is also of importance to perform
a more detailed analysis of the enveloping  algebra of $N=2$
$\widehat{sl(3)}$ with the aim to list all irreducible  composite $N=4$
supermultiplets and to study  possible $N=4$ extended  $W$ type  algebras
generated by these composite supercurrents. At last, it still remains
to classify all possible $N=2$ affine superalgebras admitting the
hidden $N=4$ structure, i.e. $N=4$ affine superalgebras. As is clear from
the two available examples ( $\widehat{sl(2)\oplus u(1)}$ and
$\widehat{sl(3)}$ ) a sufficient condition of the existence of such
a structure on the given affine superalgebra is the possibility to
define $N=4$ SCA on it via the corresponding ``long'' Sugawara
construction, with the full $N=2$ stress-tensor included.

\vskip1cm
\noindent{\Large{\bf Acknowledgments}} \\

\noindent F.T. wishes to express his gratitude to the
JINR-Bogoliubov Laboratory of Theoretical Physics, where this work
has been completed, for the kind hospitality. E.I. and S.K.
acknowledge a support from the grants RFBR-99-02-18417,
INTAS-96-0308 and INTAS-96-538.
\vspace{0.3cm}
\section*{Appendix: the second flow of the ``long'' $\widehat{sl(3)}$
$N=4$ mKdV}

\vspace{0.1cm}
For completeness we present here the evolution equations for the
second flow of the ``long''
$\widehat{sl(3)}$ mKdV hierarchy (it is the first non-trivial flow).
We have \footnote{ In order to save space and to avoid an unnecessary
duplication we present the equations only for the
non-linear chiral sector.}
\begin{eqnarray}
{\dot H} &=& - 2\partial^2 H- 2\alpha ( 2HD\partial
\overline{H}+\partial HD\overline{H}-SD\partial\overline{S}
-\partial S D\overline{S}- R D\partial \overline{R}-\partial R
D\overline{R})-\nonumber\\ && -4{\overline \alpha} \partial H
\overline{D}H + 2\alpha\overline{(F}S\partial R +
\overline{F}\partial S R - H S\partial\overline{S} - D\overline{F}
S\overline{D}R+D\overline{F}\overline{D}SR) -
\nonumber\\&&-2{\overline\alpha} HR
\partial \overline{R}- 2(1+\alpha)(H\partial S
\overline{S}+\partial H S\overline{S}) -2(1-{\overline{\alpha}}) (H\partial R
\overline{R}+\partial H R\overline{R})+\nonumber\\&&
+2(2H\overline{D}FD\overline{F}+H\overline{D}RD\overline{R}+
H\overline{D}SD\overline{S}
-2\overline{D}HFD\overline{F}-
\overline{D}HRD\overline{R}-\overline{D}HSD\overline{S}
-2H\partial F \overline{F} -\nonumber\\&&- 2\partial
HF\overline{F})
+2\alpha (2H\overline{H}FD\overline{F}+2HD\overline{H}F\overline{F}+
S\overline{S} RD\overline{R}+SD\overline{S}R\overline{R})
-\nonumber\\&&-2{\overline{\alpha}}(H\overline{H}RD
\overline{R}+HD\overline{H}R\overline{R}+
\overline{H}D\overline{F}SR
-D\overline{H}\overline{F}SR)+\nonumber\\ &&
+2(2HF\overline{S}D\overline{R}-2HF
D\overline{S}\overline{R}-H\overline{F}S\overline{D}R
+H\overline{F}\overline{D}SR
+H\overline{H}SD\overline{S}+HD\overline{H}S\overline{S}+
\overline{D}H\overline{F}SR)-\nonumber\\ &&
-2\alpha H\overline{H}\overline{F}SR -
2HS\overline{S}R\overline{R}\nonumber\\
{\dot S} &=&
2{\overline\alpha}(D\overline{H}\partial S -\overline{D}H\partial S +
D\partial \overline{H}S -\partial H\overline{D}S) -
2\overline{D}\partial H S - 2\partial FD\overline{R}-\nonumber\\
&& -2\alpha(2H\partial
\overline{H}S+SR\partial\overline{R}+S\partial
R\overline{R}+\partial H \overline{H}S)-\nonumber\\ &&
-2{\overline\alpha}
(H\overline{D}FD\overline{R}-\overline{D}HFD\overline{R}+\partial
S S\overline{S}) -\nonumber\\ && -2(F\overline{F}\partial S
+FD\overline{F}\overline{D}S + H\overline{H}\partial S
+HD\overline{H}\overline{D}S -H\partial
F\overline{R}-S\overline{D}RD\overline{R}+S\overline{D}SD\overline{S}+
2\overline{D}SRD\overline{R}+\nonumber\\ && +\partial S
R\overline{R}) +2\alpha (FSD\overline{S}R
-FS\overline{S}D\overline{R}) -2{\overline\alpha}
(HF\overline{F}\overline{D}S
+H\overline{D}HF\overline{R}+D\overline{H}F\overline{F}S+\nonumber\\
&& +\overline{H}FD\overline{F}S)
+2(1+\alpha)(\overline{F}S\overline{D}SR +
H\overline{D}SR\overline{R})-2 (
2HS\overline{D}R\overline{R}+2HS\overline{D}S\overline{S}-2H
\overline{D}F\overline{F}S-
\nonumber\\ &&
-H\overline{D}H\overline{H}S+\overline{D}HF\overline{F}S)
+2\alpha H\overline{H}F\overline{F}S
-2{\overline\alpha} H\overline{H}SR\overline{R}+2HFS\overline{S}R\nonumber\\
{\dot R} &=& 2\alpha (\overline{D}\partial H R -
D\overline{H}\partial R) -2{\overline\alpha}(\overline{D}H\partial R
+\partial H \overline{D}R) -2(D\partial \overline{H}R -\partial
FD\overline{S}) +\nonumber\\ && +2\alpha (H\partial
F\overline{S}-\partial R R\overline{R}) +2{\overline\alpha}
(H\overline{D}FD\overline{S}-2H\partial\overline{H}R-S\partial\overline{S}R-
\overline{D}HFD\overline{S}-\nonumber\\ && -\partial
H\overline{H}R -\partial S\overline{S}R)-2(F\overline{F}\partial R
+FD\overline{F}\overline{D}R + H\overline{H}\partial R +H
D\overline{H}\overline{D}R +R\overline{D}RD\overline{R} +S
\overline{S}\partial R +\nonumber\\ &&+ 2
SD\overline{S}\overline{D}R-\overline{D}SD\overline{S}R ) +
2\alpha(2HR\overline{D}R
\overline{R}-2H\overline{D}F\overline{F}R-H\overline{D}H
\overline{H}R-2H\overline{D}S \overline{S}R +\nonumber\\
&&+\overline{D}HF\overline{F}R) + 2{\overline\alpha}
(F\overline{S}RD\overline{R}+FD\overline{S}R\overline{R}-HF
\overline{F}\overline{D}R)
+2(1+{\overline\alpha})(\overline{F}SR\overline{D}R) -\nonumber\\
&&-2(1+\alpha )HS\overline{S}\overline{D}R
-2(H\overline{D}HF\overline{S}-\overline{H}FD\overline{F}R
-D\overline{H}F\overline{F}R) +\nonumber\\ &&+2\alpha
(HF\overline{S}R\overline{R}-H\overline{H}S\overline{S}R)
+2{\overline\alpha} H\overline{H}F\overline{F}R \nonumber\\ {\dot F}
&=& 2\partial^2 F -4{\alpha}D\overline{H}\partial F
-4{\overline\alpha}( \overline{D}H\partial F +\overline{D}\partial H
F) +2(\overline{D}S\partial R -S\overline{D}\partial R
+\overline{D}\partial S R-\nonumber\\ &&
 -\partial S\overline{D}R)-2{\overline\alpha} (4HF\overline{D}F\overline{F}
 -2H\overline{D}H\overline{H}F+\overline{H}FRD\overline{R}-
D\overline{H}FR\overline{R})
 -\nonumber\\
 &&
 -2\alpha\overline{D}HFS\overline{S}-2(1+{\overline\alpha})
(H\overline{D}FS\overline{S}+
 HF\overline{D}S\overline{S})
 +2i\sqrt{3}(HF\overline{D}R\overline{R}+H\overline{D}FR\overline{R})+
\nonumber\\
 && +2 (2F\overline{F}S\overline{D}R -2F\overline{F}\overline{D}SR
 +H\overline{H}S\overline{D}R -H\overline{H}\overline{D}SR +\overline{H}
FSD\overline{S}+2SR\overline{D}R\overline{R}-2S\overline{D}S\overline{S}R
+\nonumber\\ && +2\overline{D}F\overline{F}SR +\overline{D}H FR
\overline{R}+\overline{D}H\overline{H}SR
-D\overline{H}FS\overline{S})+\nonumber\\ &&+2\alpha
(D\overline{H}S\overline{D}R-D\overline{H}\overline{D}SR) -
2{\overline\alpha}\partial \overline{H}SR
-2(FR\partial\overline{R}+FS\partial\overline{S}+2F\overline{D}FD
\overline{F}
+\nonumber\\ && +F\overline{D}RD\overline{R}+
F\overline{D}SD\overline{S}+2H\overline{H}\partial F + 2 HD
\overline{H}\overline{D}F +2H\partial\overline{H}F +\overline{D}FR
D\overline{R}+\overline{D}FSD\overline{S}+\nonumber\\&&+ 2\partial
F F\overline{F}+2\partial F R\overline{R}+2\partial F
S\overline{S}) +2(1+\alpha) H\overline{H}FR\overline{R}
+2(1+{\overline\alpha}) H \overline{H}FS\overline{S}~.\nonumber
\end{eqnarray}
The parameters $\alpha, \bar\alpha$ have been defined in eq. \p{param}.

\end{document}